\documentclass[%
 reprint,
superscriptaddress,
bibnotes,
 amsmath,amssymb,
prapplied,
]{revtex4-1}
\usepackage{graphicx}
\usepackage{dcolumn}
\usepackage{bm}
\usepackage{hyperref}
\usepackage{natbib}
\usepackage{subfigure}
\usepackage{tikz}
\usepackage{pgfplots}
\usetikzlibrary{decorations.markings}
\pgfplotsset{width=10cm,compat=1.9}
\pgfplotsset{yticklabel style={text width=2em,align=right}}
\begin{document}


\title{Mechanical transmission of rotational motion between molecular-scale gears}

\author{H.-H. Lin}
 \email{hhlin@nano.tu-dresden.de}
\affiliation{%
  Institute for Materials Science and Max Bergmann Center of Biomaterials, TU Dresden, 01069 Dresden, Germany\\
}
\affiliation{
Max Planck Institute for the Physics of Complex Systems, 01187 Dresden, Germany\\
}
\author{A. Croy}%
\affiliation{%
  Institute for Materials Science and Max Bergmann Center of Biomaterials, TU Dresden, 01069 Dresden, Germany\\
}
\author{R. Gutierrez}
\affiliation{%
  Institute for Materials Science and Max Bergmann Center of Biomaterials, TU Dresden, 01069 Dresden, Germany\\
}
\author{C. Joachim}
\affiliation{%
 GNS and MANA Satellite, CEMES-CNRS, 29 rue J. Marvig, 31055 Toulouse Cedex, France\\
}
\author{G. Cuniberti}
\homepage{https://nano.tu-dresden.de/}
\affiliation{%
  Institute for Materials Science and Max Bergmann Center of Biomaterials, TU Dresden, 01069 Dresden, Germany\\
}
\affiliation{%
Dresden Center for Computational Materials Science, TU Dresden, 01062 Dresden, Germany\\
}
\affiliation{%
Center for Advancing Electronics Dresden, TU Dresden, 01062 Dresden, Germany\\
}

\date{\today}

\begin{abstract}
Manipulating and coupling molecule gears is the first step towards realizing molecular-scale mechanical machines. Here, we theoretically investigate
the behavior of such gears using molecular dynamics simulations. Within a nearly rigid-body approximation we reduce the dynamics of the gears to the rotational motion around the orientation vector. This allows us to study their behavior based on a few collective variables. Specifically, for a single hexa (4-tert-butylphenyl) benzene molecule we show that the rotational-angle dynamics corresponds to the one of a Brownian rotor. For two such coupled gears, we extract the effective interaction potential and find that it is strongly dependent on the center of mass distance. Finally, we study the collective motion of a train of gears. We demonstrate the existence of three different regimes depending on the magnitude of the driving-torque of the first gear: underdriving, driving and overdriving, which correspond, respectively, to no collective rotation, collective rotation and only single gear rotation. This behavior can be understood in terms of a simplified interaction potential. 
\end{abstract}

\maketitle


\section{\label{sec:Introduction}Introduction}
Miniaturizing gears towards the molecular-scale size\cite{Soong2000,Yang2014} or using a
molecule to function like a nanogear\cite{Manzano2009,Michl2009} opens new paths for the construction of
nanoscale mechanical machinery\cite{Deng2011,Chiaravalloti2007}. A large number of experiments have been performed
to address the issue of driving the rotation of a single-molecule rotor on a surface by the
scanning tunneling microscope (STM)\cite{Gimzewski1998,Tierney2011,Ohmann2015,Perera2013,Gao2008}. From the
theoretical side, there have also been several proposals for inducing the rotation of a single molecule-
rotor mounted on an axle, such as quantum rotors\cite{Bustos-Marun2013,Bruch2018,Bustos-Marun2019,Liu2019} or tunneling current-induced
rotations\cite{PES,Wang2008,Croy2012,Lin2019,Wachtler2019}. However, transferring angular momentum mechanically from one gear to the
next in a train of molecule-gears is a very challenging problem from the experimental point of view \cite{Gao2008,Mishra2015,WeiHyo2019}. Although mechanical transmission involving up to three molecule-gears\cite{WeiHyo2019} as well as 
collective rotations have recently been observed\cite{Chao,Zhang2016}, it is still unclear
how to design single molecule-gears in order to control  mechanical rotations along a long train of
molecule-gears. Several calculations using density functional theory (DFT) have been
performed trying to establish specific design rules concerning, for example, the axle stability or
the gear teeth flexibility\cite{Hove2018,Hove2018a, Zhao2018, Hove2018a,Chen2018}. However, there are many open questions about how molecule-gears must be individually stabilized on a surface and how they must mutually interact 
in a long train of gears for the mechanical transmission of motion to occur along the train.
\begin{figure}[!h]\centering
 \includegraphics[width=0.5\textwidth]{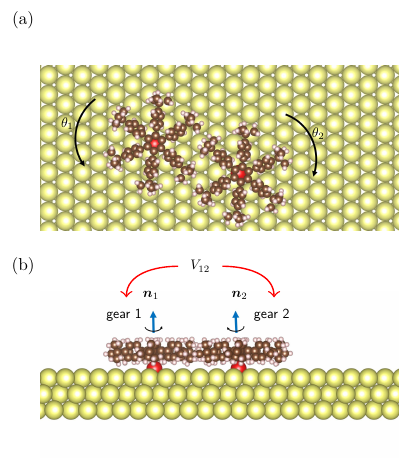}
 \caption{Schematic plot of a train of molecule gears with two hexa (4-tert-butylphenyl)benzene mounted above copper atoms (red) on top of a lead surface (yellow). (a) Top view with rotational angle $\theta_1$ and $\theta_2$. (b) Side view with rotational axes $\boldsymbol{n}_1$ and $\boldsymbol{n}_2$.
  }
 \label{Fig:Scheme}
\end{figure}

In experiments, the molecule-gears are either chemically or physically adsorbed on the surface. They are mounted on their rotational axle and mutually interact via van-der-Waals forces, hydrogen bonds or dipole-dipole interactions -- depending on the nature of the molecules. In this article, we consider the scenario shown in Fig.~\ref{Fig:Scheme} where the gears are realized by identical molecules. To be specific, we chose hexa (4-tert-butylphenyl)benzene physically mounted above copper atoms (red) on top of a lead (yellow) surface. This setup has been recently  shown to be a successful platform for implementing a train of molecule-gears\cite{WeiHyo2019}. 

Depending on the specific setup, different computational methods are available to describe the molecule-gears. For instance, DFT has been used to study a 5-membered carbon (cyclopentadienyl) ring with cyano group mounted on a manganese atom above a graphene surface\cite{Hove2018} and to investigate PF$_3$ molecules on a Cu(111) surface\cite{Hove2018a}. Such calculations are computationally demanding and the system sizes and time-scales which can be realized are limited when compared to experimentally relevant scales. For physisorbed gears, on the other hand, classical molecular dynamics (MD) provides a suitable approach. Similarly, transmission between gears based on carbon nanotubes\cite{Han1997} and fullerenes,\cite{Robertson1994} but without supporting surface, have been studied using MD. Additionally, the qualitative behavior of coupled gears can be described in model-based approaches, where the degrees of freedom of each gear are reduced to their respective rotational angle. Here, the form of the effective inter-gear potential plays a crucial role for the ability to transmit rotations along a train of such gears.

This paper is organized as follows: In section \ref{sec:Formalism}, we first introduce a nearly rigid-body approximation to reduce the number of degrees of freedom and to define the gear orientation-vector. We then treat individual gears as Brownian rotors and review some properties for the single gear and a train of gears, respectively. In the latter case, we use a probabilistic approach to extract the interaction potential between gears from MD simulations. In section \ref{sec:Applications}, we apply our methods to a molecule-gear made of hexa (4-tert-butylphenyl)benzene and discuss the general properties in thermal equilibrium and the rotational behavior under external torque. In the case of a train of gears, the effect of the center-of-mass distance between gears is also studied. 

\section{\label{sec:Formalism}Modelling and methodology}
We consider the generic setup shown in Fig.~\ref{Fig:Scheme} consisting of two molecule-gears (for simplicity, we will just call them gears from now on) on a Pb(111) surface\cite{WeiHyo2019}. The center-of-mass of each gear is anchored to a copper atom such that the gear rotation axis is always fixed. We further assume the distance between gears and substrate to be sufficiently large such that the gear-substrate interaction potential mainly depends on the vertical distance $d$ to the substrate, but only weakly on the exact horizontal location of the gear. Thus, we expect that there is an optimal distance confining the rotation of gears to be parallel to the substrate (i.e.\ confined to the $xy$-plane). The gear-substrate interaction  can be thus represented by a Lenard-Jones type of potential. Finally, the gears are considered to be in thermal equilibrium with their environment, i.e.\  with the substrate, so that any energy dissipation effects arise from energy relaxation into the substrate. 

\subsection{\label{sec:MD}Molecular dynamics}
In order to describe the previously described experimental situation, we use MD simulations. The gears
are assumed to be weakly coupled to the metal substrate, so that no electron transfer needs to be included. Based on our previous assumptions, the interaction with the substrate is only dependent on the molecule-surface distance and we will not consider the underlying substrate at the atomic level. Rather, an artificial substrate is considered, whose only effect on the molecule(s) is described via a 9-3 Lennard-Jones potential {with parameters $\epsilon=0.1$~eV and $\sigma=5$~\AA} applied to all atoms of the molecules. Each gear is placed at a distance $d=5$~\AA{} from the substrate and its center-of-mass is fixed. For the interatomic potential, we use the adaptive intermolecular reactive empirical bond-order (AIREBO) potential\cite{Stuart2013}, which works well for most hydrocarbon materials. We use a Langevin thermostat with relaxation time $\tau=1$ ps\cite{Persson1996} as implemented in LAMMPS\cite{Plimpton1995}. Finally, in order to evaluate the rotational dynamics, we need to define the orientation-vector of the gear which is a collective variable including the motion of all atoms.

\subsection{\label{sec:NRBA}Nearly rigid-body approximation}
We consider a set of coupled gears, characterized by a set of Cartesian vectors $\lbrace \boldsymbol{r}_{\alpha k}\rbrace$ with $\alpha$ labelling the gears and $k$ running over all atoms in gear $\alpha$, respectively. In principle, the trajectory $\lbrace \boldsymbol{r}_{\alpha  k}(t)\rbrace$ can be separated into rigid-body motion, which entails the center-of-mass and the rotational degrees of freedom, and internal motion\cite{Littlejohn1997,PES}. By choosing a reference structure $\lbrace \boldsymbol{r}_{\alpha  k}^{(0)}\rbrace$, we can always find a unique set of angles $\theta_{\alpha}$ and rotation axes $\boldsymbol{n}_{\alpha}$.
Using rotation matrices $\lbrace\mathbf{R}(\theta_{\alpha}, \boldsymbol{n}_{\alpha} )\rbrace$ with respect to the center-of-mass position the total deviation of all coordinates from the translated and rotated reference coordinates can be minimized. Concretely, the error associated with rotation angle $\theta_{\alpha }$ for the $k^{th}$ atom of the $\alpha ^{th}$ gear is defined as:
\begin{equation}
	\boldsymbol{\epsilon}_{\alpha k}=\mathbf{R}(\theta_{\alpha}, \boldsymbol{n}_{\alpha}) \boldsymbol{r}_{\alpha k}-\boldsymbol{r}_{\alpha k}^{(0)}\;.
\end{equation}
Then we aim at minimizing the weighted sum of the errors squared for each gear:
\begin{equation}
	\epsilon^{\alpha}_{tot}=\sum_{k}w_{{\alpha} k}|\boldsymbol{\epsilon}_{{\alpha} k}|^2\;,
\end{equation}
where the weight $w_{{\alpha}k}=m_{{\alpha}k}/M_{\alpha}$ is the ratio between mass $m_{{\alpha}k}$ of the $k^{th}$ atom of the ${\alpha}^{th}$ gear and total mass $M_{\alpha}$. Technically, this can be efficiently implemented by using quaternions.\cite{Kneller1991} As a result, the molecular orientation-vector and the respective rotation angle along the trajectory can be extracted. We have thus reduced a large number of degrees of freedom to only a few relevant variables. We call this reduction method {nearly rigid-body approximation}, since its validity is limited to gears displaying fairly small deformations. In this framework, the degrees of freedom of each individual gear are reduced to a single variable $\theta_{\alpha}$ and the corresponding axis of rotation $\boldsymbol{n}_{\alpha}$.

\subsection{\label{sec:analytic}Analytic method}
The nearly rigid-body approximation can be applied to reduce the number of degrees of freedom in each gear, so that we can consider the rotational dynamics directly for these collective variables, only. This is of particular advantage when treating the rotational motion using generic models, as we illustrate in the next subsections.

\subsubsection{\label{sec:brownian}Single gear}
First, we consider the random rotation of a single gear under the influence of thermal fluctuations due to the surface (no net external driving). The simplest way to account for this situation is a free Brownian rotor, which is described by the {Langevin equation}\cite{Risken1989}:
\begin{equation}
    I\ddot{\theta}=-\gamma\dot{\theta}+\xi(t)\;.
\end{equation}
Here, $I$ is the moment of inertia of the molecule with respect to the rotation axis (e.g.\ the $z$-axis), $\gamma$ is a damping coefficient and $\xi(t)$ is a stochastic torque, given by Gaussian white noise:
\begin{equation}
    \langle \xi(t_1)\xi(t_2)\rangle=g\delta(t_1-t_2)\;,
\end{equation}
 $g$  denoting the strength of the time correlation. In thermal equilibrium, the fluctuation-dissipation theorem holds and one obtains $g=2\gamma k_B T$, where 
$k_B$ is the Boltzmann constant and $T$ is the temperature of the surface. An important result of free Brownian rotation is that the long time limit ($\gg \gamma^{-1}$) of the variance of $\theta$ satisfies the following linear dependence on time\cite{Kampen2007}:
\begin{equation}\label{eq:slope}
    \langle \theta^2(t)\rangle=\frac{2 k_B T}{\gamma} t\;.
\end{equation}
This relation is typical for a diffusion process and, in the context of gears, has also been observed in experiments\cite{Gao2008,Gimzewski1998,Tierney2011,Perera2013}. In Sec.\ \ref{sec:App_Single} this formula will be used to verify the Brownian rotation of a single gear in the MD simulations.

\subsubsection{\label{subsec:gears}Many gears}
Consider next a linear train of gears. We assume all gears are the same and they are coupled to their nearest neighbors via a short-range interaction, which can be written as a superposition of Lenard-Jones potentials. Then we can easily write a Langevin equation for the $i^{th}$ gear as follows: 
\begin{eqnarray}
    I_i\ddot{\theta}_i=-\gamma\dot{\theta}_i-\frac{\partial}{\partial \theta_i}\left( V_{i,i-1}+ V_{i,i+1}\right)+\xi_{i}(t)\;,
\end{eqnarray}
where $V_{i,i-1}$ and $V_{i,i+1}$ are the two-gear interaction potentials. In general, they can be written as a function of the gear angle and the distance $d_{i,i+1}$ between gears, namely:
\begin{equation}
    V_{i,i+1}=V_{i,i+1}(\theta_i,\theta_{i+1},d)\;.
    \label{eq:V_12}
\end{equation}
In order for the total angular momentum to be conserved in absence of noise, the two-gear interaction potential has to adopt the following generic form\cite{MacKinnon2002,Mackinnon2005}:
\begin{equation}
    V_{i,i+1}(\theta_i,\theta_{i+1},d)=V_0 u(\theta_i+\theta_{i+1}) \quad (d<d_0)\;,
    \label{eq:V_ideal}
\end{equation}
where $d_0$ is the largest distance where two gears are still interacting and $u(\theta)$ is a periodic function of period $2\pi/N_{\text{teeth}}$ ($N_{\text{teeth}}$ is the  number of teeth) with a single minimum at $u(0)=0$. The amplitude $V_0$ determines the softness of the gears. For instance, when $V_0$ approaches infinity, this requires $u(0)=0$ or $\theta_i=-\theta_{i+1}$, which means the gears are rigid bodies and perfectly interlocked.

\section{\label{sec:Applications}Results: planar rotation of gears} 
In this section, we show MD simulation results for single and few interacting gears in thermal equilibrium with the surface as well as for the case where one gear is driven by an external torque. We compare our simulations with the results obtained using the Langevin equation approach presented in section \ref{sec:analytic}.

\subsection{Single gear\label{sec:App_Single}}

\begin{figure}[!tb]\centering
 \includegraphics[width=0.45\textwidth]{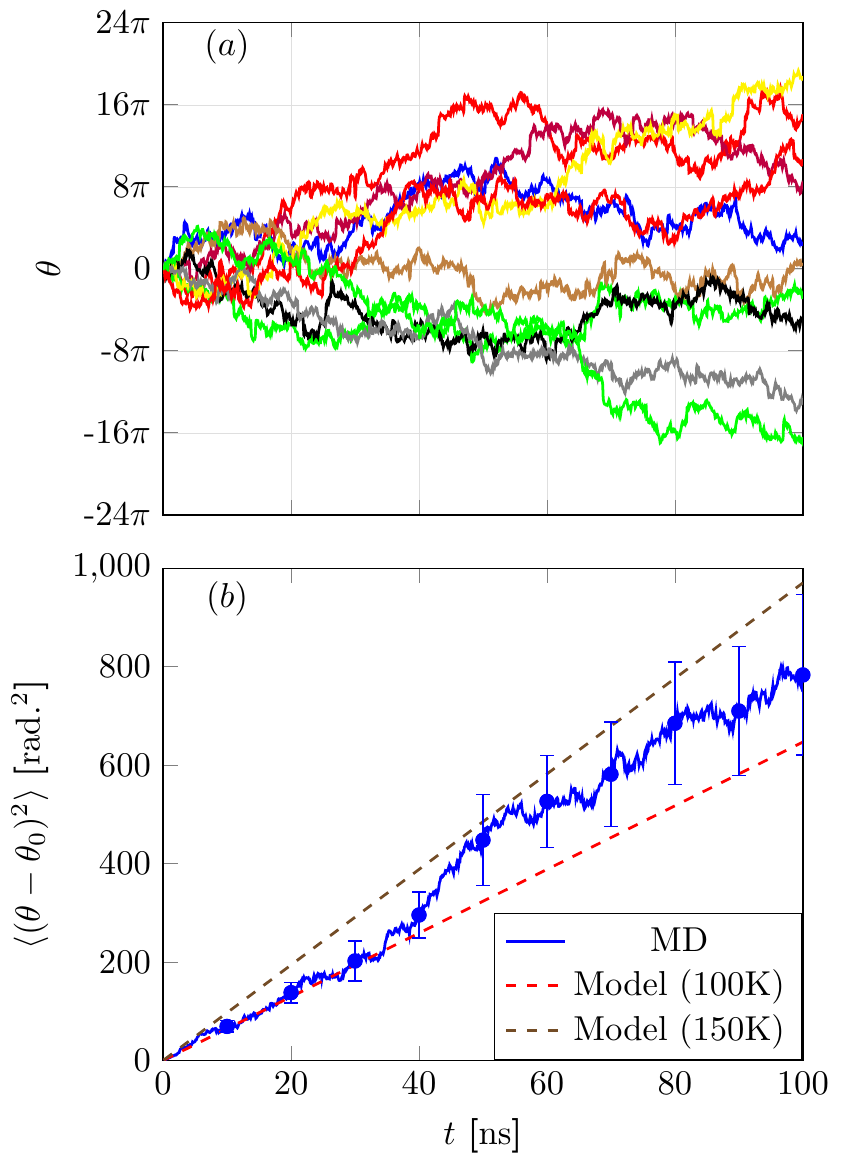}
 \caption{(a) Trajectories of gear orientation $\theta(t)$ at $T=100$ K from $10$ different random noise realizations (distinguished by colors), which shows a typical 1D random walk pattern. (b) Ensemble average of angle variance $\langle \theta^2\rangle$ from MD simulations (in blue dots) of $50$ different trajectories with standard error of mean; the dashed line shows the theoretical value of variance from the model of 1D Brownian rotation, which is linearly proportional to time $\langle \theta^2\rangle=(2 k_B T \tau/I )t$, where the temperatures $T=100$ and $T=150$ K, relaxation time $\tau=1$ ps and molecule moment of inertia {$I=2.13\times 10^{-41}$kg$\cdot$ m$^2$}.}  
 \label{Fig:Fig2}
\end{figure}

In order to facilitate the sampling of statistically relevant trajectories $\theta(t)$ during the MD run, we first consider the case where the temperature is $T=100$ K. In Fig.~\ref{Fig:Fig2}~(a), we plot the trajectories $\theta(t)$ from $10$ different random noise realizations. As one can see, the trajectories display a typical 1D random walk pattern with increasing variance over time (the longer the time elapses, the wider the variance becomes). In Fig.~\ref{Fig:Fig2}~(b), we plot the ensemble average of the angles (blue curve) from $50$ different trajectories together with the standard error of the mean. 
This curve can be compared to the analytical result obtained from the free Brownian rotor given by Eq.\ \eqref{eq:slope} using
for the relaxation time $\tau=1$ ps and for the molecule moment of inertia  {$I=2.13\times 10^{-41}$kg$\cdot$ m$^2$} (by taking the component of the moment of inertia tensor associated to $z$-axis rotations calculated within LAMMPS).
The red dashed line indicates the resulting behavior when the temperature is set to the simulation temperature ($T=100$ K). However, one observes a non-negligible deviation between the model and the MD result. This can be traced back to the freezing of displacement degrees of freedom for motion in z-direction due to the presence of the substrate. Hence, according to the equipartition theorem, the effective temperature of the molecule will be larger than the value used for the thermostat. Using a temperature of $150$ K in the analytical model yields a closer agreement with the MD result, as shown in Fig.~\ref{Fig:Fig2}~(b) (brown dashed line). We further find that the axis of rotation remains almost parallel to the $z$-axis during the simulation time.

\begin{figure}[!tb]\centering
 \includegraphics[width=0.4\textwidth]{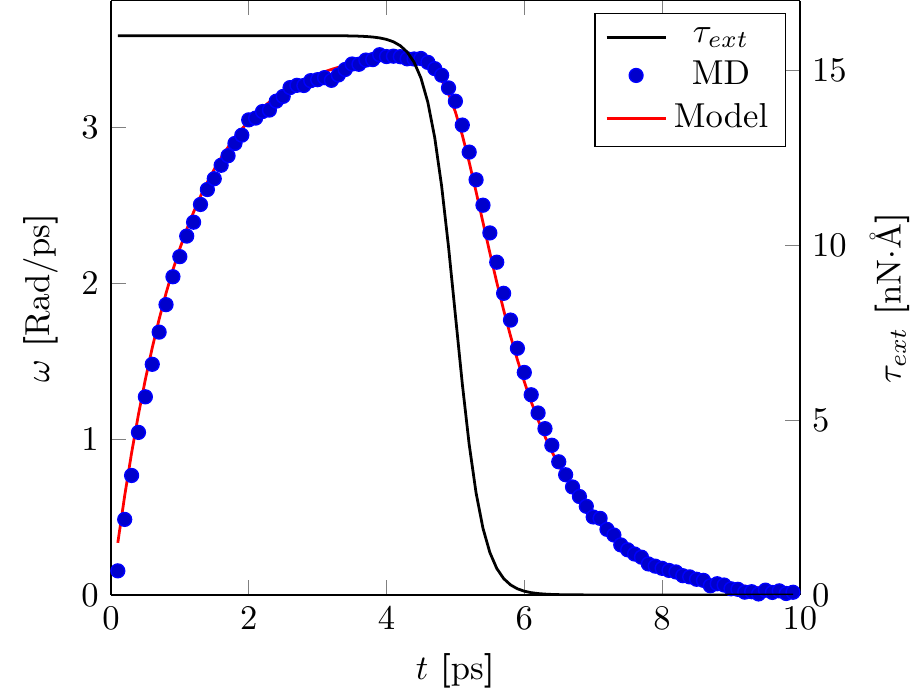}
 \caption{Plot of angular velocity under external time-dependent torque (in black curve) for the MD simulation (blue symbols) in comparison to the result from the model (in red line). Here we use the parameters $T=10$ K, relaxation time $\tau=1$ ps and molecule moment of inertia $I=2.13\times 10^{-41}$kg$\cdot$ m$^2$.}  
 \label{Fig:Fig3}
\end{figure}

In the MD simulations, we now apply an external torque to drive the gear and to control the rotational directionality. As seen above, at a high temperature ($T=100$ K) stochastic rotations are dominant. We therefore lower the temperature to $T=10$ K to make sure that only driven rotations dominate on the time-scale of nanoseconds. In Fig.~\ref{Fig:Fig3}, we plot the angular velocity $\omega=\dot{\theta}$ from the MD simulations (blue line) under an external time-dependent torque $\tau_{ext}$ (black line), which is switched off from and initial value of $16$ nN$\cdot$\AA{}. In comparison, we show the solution for an ideal gear with suppressed fluctuations, which satisfies:
\begin{equation}
    I\dot{\omega}(t)=-\gamma \omega(t) + \tau_{ext}(t)\;,
\end{equation}
where $\gamma=I/\tau$ is the damping coefficient. Again, we use the relaxation time $\tau=1$ ps and the same molecule moment of inertia as above. One can see that the angular velocity obtained from the model is almost identical to the one from MD. Thus, the rigid-body equation of motion provides a reasonable description for the gear since internal degrees of freedom do not considerably affect the moment of inertia. Additionally, this approach allows us to demonstrate that the variable $\theta$ can effectively emerge out of the multiple excitation of individual atoms in the MD simulation.   

\subsection{A train of gears}
We address now a train of two gears with the goal of clarifying under which conditions the transmission of rotations is possible and discuss the influence of the center-of-mass distance in thermal equilibrium. Then, we apply an external torque to drive one of the gears to investigate if rotation still occurs. 

\subsubsection{Thermal equilibrium}
 \begin{figure*}[htb]\centering
 \includegraphics[width=1.1\textwidth]{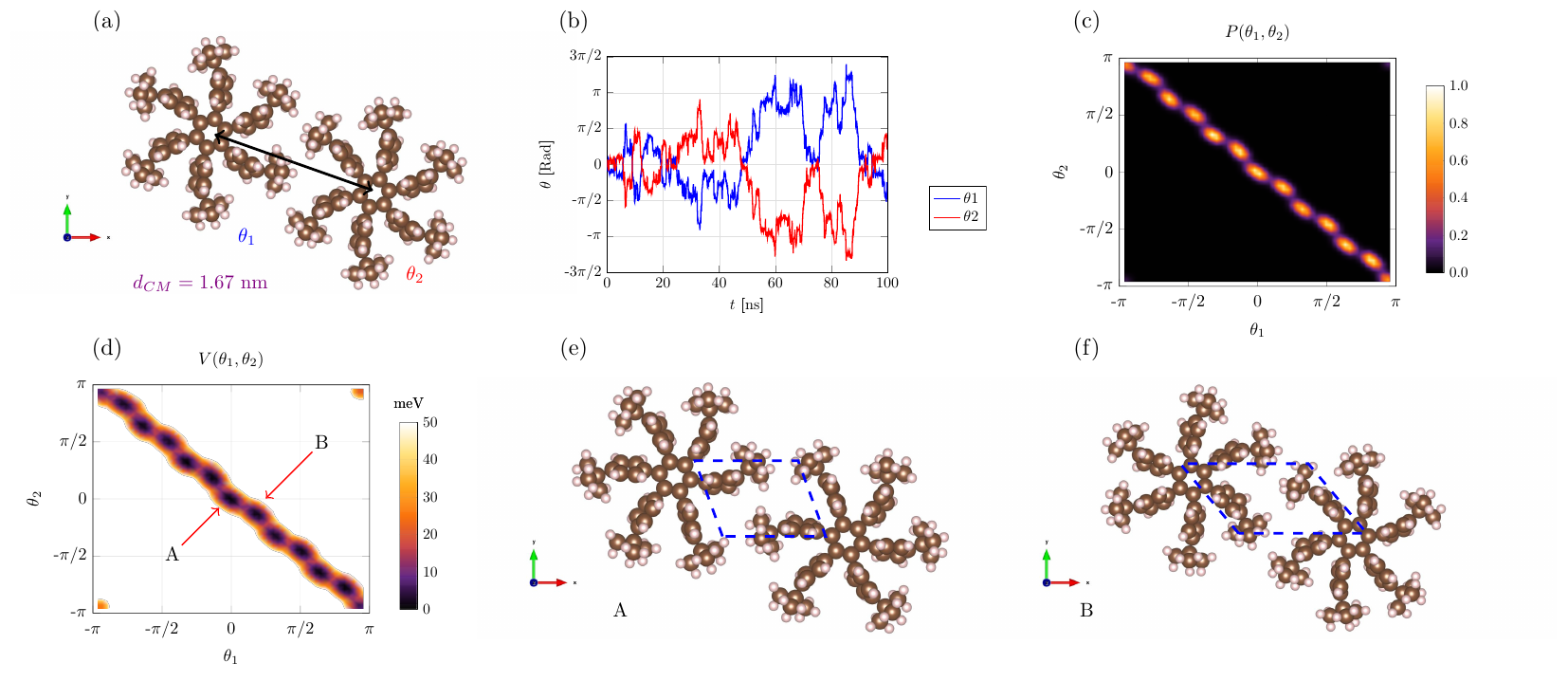}
 \caption{(a) Schematic plot of two coupled hexa(4-tert-butylphenyl)benzene gears with center-of-mass distance $d_{CM}=1.67$nm. (b) The molecular orientations $\theta_1$ and $\theta_2$ correspond to the left and the right gear, respectively. The trajectories manifest a collective rotation in thermal equilibrium at $T=100K$. (c) Joint-probability distribution $P(\theta_1,\theta_2)$ of orientations $\theta_1$ and $\theta_2$ from an ensemble of $N=50$ pairs of gears. (d) Two-gear interaction potential $V(\theta_1,\theta_2)=-k_BT\text{ln}P(\theta_1,\theta_2)$, which shows $12$ local minima along the diagonal with inner barrier heights $V_{A-B}\approx 10$ meV. (e)-(f) Metastable states at local minima A and B, respectively, shown in (d) with tert-butyl groups anti-parallel from top to down and left to right.}  
 \label{Fig:Fig4}
\end{figure*}
Following the experimental setup\cite{WeiHyo2019}, the molecular structure in Fig.~\ref{Fig:Fig4}~(a) (see also Fig.~\ref{Fig:Scheme}~(a)) will be used as a reference (to define the orientation of a gear) and as initial conformation in the MD simulations. The center-of-mass distance is taken as $d_{CM}=1.67$ nm. First, we perform the structural optimization and then switch-on the temperature ($T=100$ K) to evolve the  system (using Langevin dynamics for $100$~ns.). As shown in Fig.~\ref{Fig:Fig4}~(b), the trajectories include both stochastic and concerted rotations. Also, the orientation $\theta_2$ is the negative of $\theta_1$ ($\theta_2\approx-\theta_1$), indicating that the two gears are {interlocked}. Notice also that the amplitude of the fluctuations of the rotations is now smaller than for the single-gear case.

Furthermore, the trajectories provide a way to extract the interaction potential $V(\theta_1,\theta_2)$ between the two gears. {As in the case of one gear (cf.\ Fig.~\ref{Fig:Fig2}), the two gears will be kicked by the random noise at finite temperatures and one obtains a distribution of angles $(\theta_1, \theta_2)$ considering an ensemble of $50$ trajectories and all time-steps. Then one can use a Gaussian kernel-density estimation\cite{Silverman1986} to obtain the probability density function $P(\theta_1,\theta_2)$. The result is shown in Fig.~\ref{Fig:Fig4}~(c)}. One can see that the main contribution is coming from the diagonal $\theta_1+\theta_2=0$, which is qualitatively consistent with perfectly interlocked gears. To rationalize this, we convert the obtained probability distribution $P(\theta_1,\theta_2)$ to the two-gear interaction potential $V(\theta_1,\theta_2)$ {by inverting the Boltzmann distribution:
\begin{equation}
 V_{12}(\theta_1,\theta_2)=-k_B T \text{ln}P(\theta_1,\theta_2)+\text{const.}\;,
 \label{eq:Boltzmann}
\end{equation}}
thus obtaining the function shown in Fig.~\ref{Fig:Fig4}~(d). In this panel, one sees that for every $30^{\circ}$ rotation there is a local minimum ($12$ minima in total) with neighboring potential barriers of about $10$~meV, the local minima corresponding to metastable states. This value is smaller than the energy necessary to dismount the gear from its axle. To shed additional light, we consider the local minima at point A and B in Fig.~\ref{Fig:Fig4}~(d) and extract the corresponding snapshots from the MD trajectories. As shown in Fig.~\ref{Fig:Fig4}~(e) and (f), the gears will interlock in two different ways: the tert-butyl groups can be anti-parallel either from top to down or left to right and this pattern will interchange every $30^{\circ}$. One thing worth mentioning is that the local minima are not exactly along the diagonal but are slightly shifted either up or down, depending on the metastable states. This means that the gear orientation can have a relative phase difference $\theta_1+\theta_2=\pm\Delta\phi$, since the teeth (or tert-butyl groups) are not stiff enough to fix their position.        
\begin{figure*}[tb]\centering
 \includegraphics[width=0.95\textwidth]{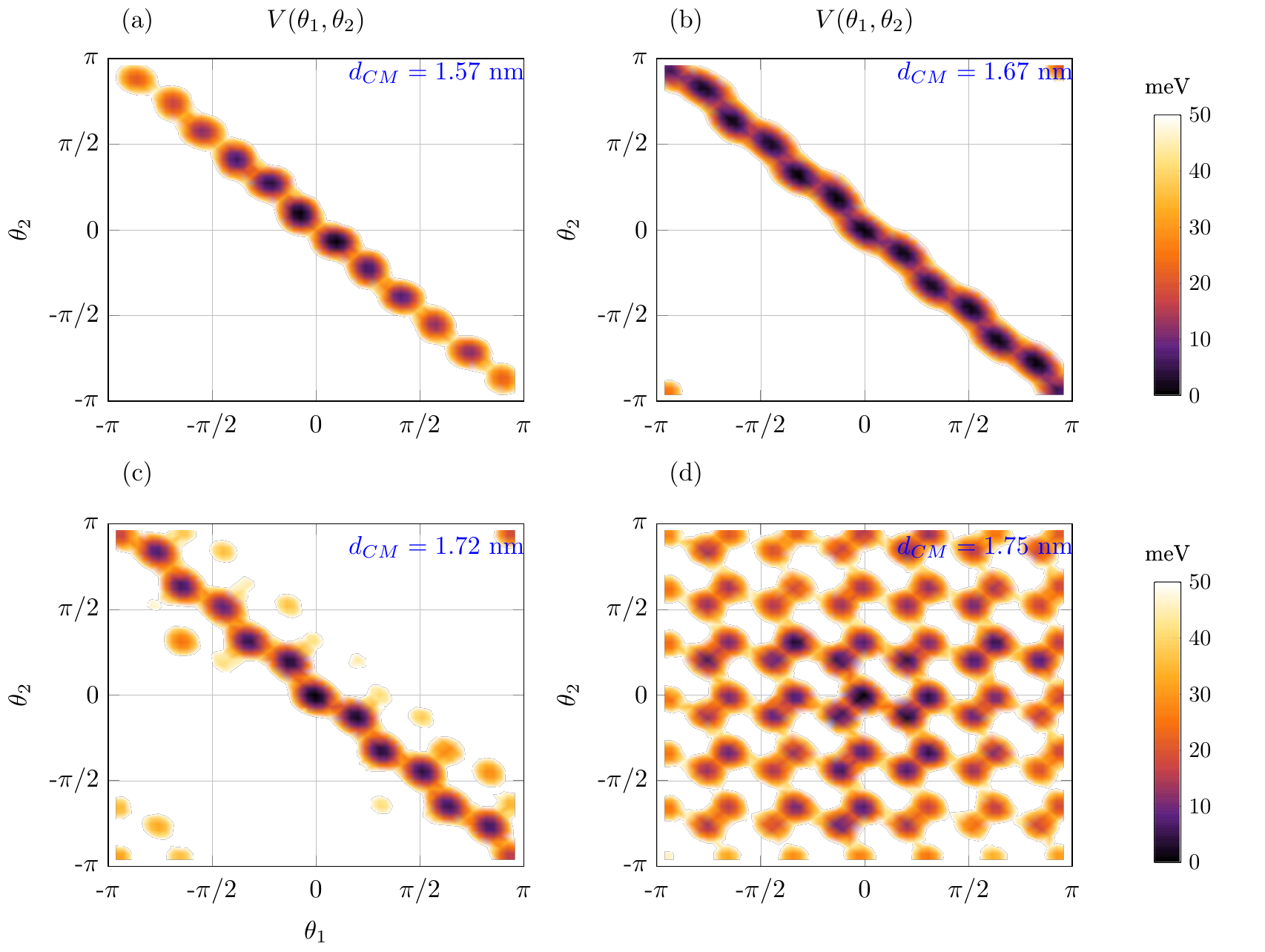}
 \caption{Two-gear interaction potential for different center-of-mass distances (a) $d_{CM}=1.57$ nm, (b) $1.67$ nm, (c) $1.72$ nm and (d) $1.75$ nm, respectively.}  
 \label{Fig:Fig5}
\end{figure*}

So far the discussion is based on one center-of-mass distance $d_{CM}=1.67$ nm, but it is obvious that the intermolecular distance plays an important role for building gears systems. In Fig.~\ref{Fig:Fig5}, we show the two-gear interaction potential with respect to several different center-of-mass distances. One can see in Fig.~\ref{Fig:Fig5}\ (a) that all local minima become very much aligned along the diagonal given by $\theta_1+\theta_2=0$ for $d_{CM}=1.57$ nm. This can be understood as follows: when the two gears become much closer, the tert-butyl groups of one of the gears will reach deeper into the core of the other one. Since the inner phenyl groups are less deformable than the outer tert-butyl groups, the freedom for teeth deformation will be suppressed and, hence, the relative phase difference $\Delta \phi \approx 0$ will be reduced. On the contrary, if we increase the distance to $d_{CM}=1.72$ nm (see Fig.~\ref{Fig:Fig5}\ (c)), then the phase difference $\Delta \phi$ will in turn increase. Also, one can see that some tiny islands appear near the main diagonal, which  are due to the small probability for the gears to have \textit{full-step} phase difference or $\Delta \phi \rightarrow \Delta \phi \pm 30^{\circ}$. Moreover, if we increase the distance even further ($d_{CM}=1.75$ nm, see Fig.~\ref{Fig:Fig5}\ (d)), then a complicated pattern is found and the two gears are no longer interlocked allowing multiple full-steps phase differences $\Delta \phi \rightarrow \Delta \phi \pm n\times 30^{\circ}$, where $n$ is an integer.

\subsubsection{External torque}\label{sec:drv_gears}
\begin{figure*}[!tb]\centering
 \includegraphics[width=0.9\textwidth]{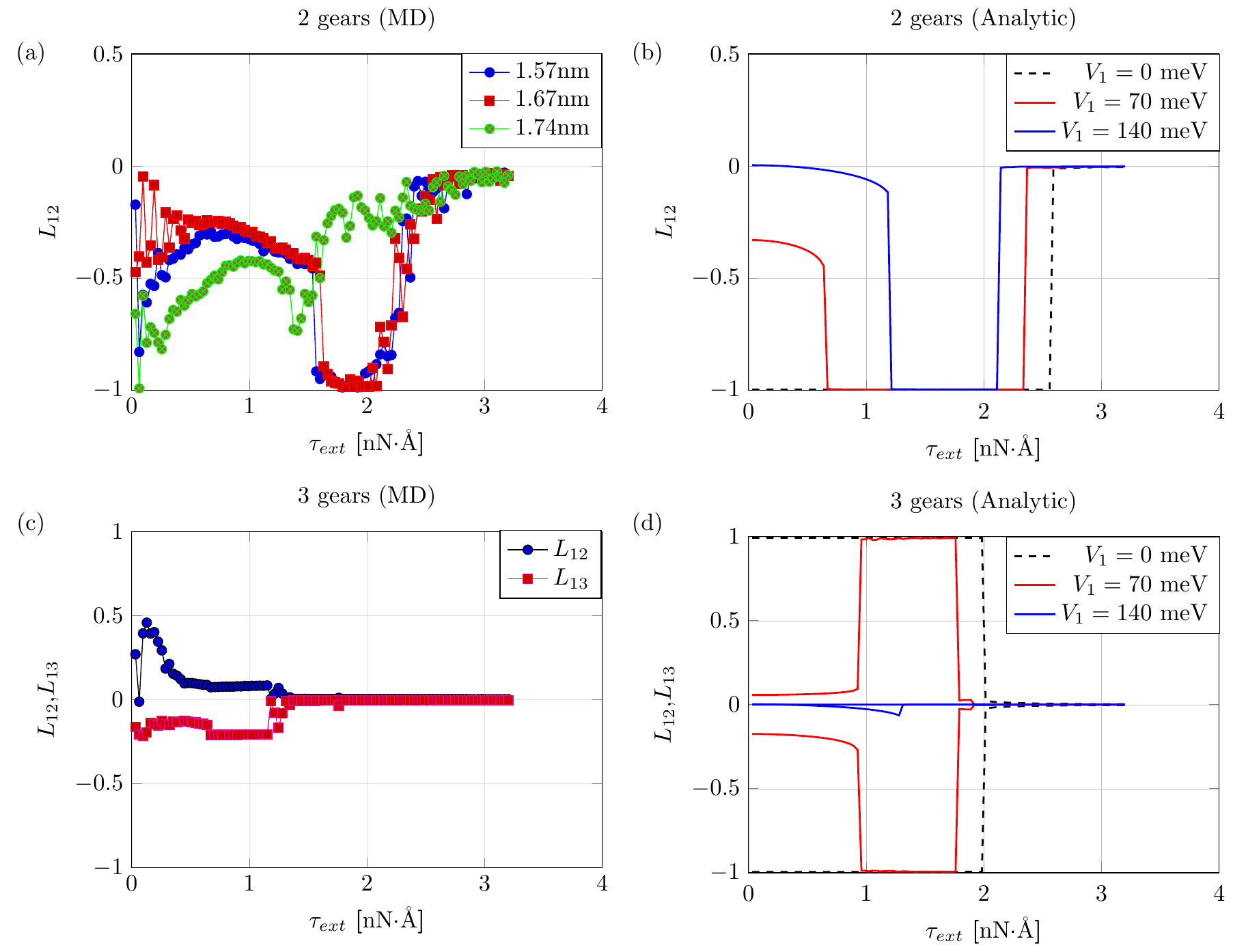}
 \caption{(a) Locking coefficient $L_{12}$ as given by Eq.\ \eqref{eq:locking_coeff} and with simulation time $T_s=500$ ps as a function of external torque $\tau_{ext}$ for different center-of-mass distances $d_{CM}=1.57$, $1.67$ and $1.74$ nm.  (b) Two gears locking coefficient from the analytic two-gear potential for different barrier heights $V_1=0$, $70$ meV and $140$ meV. (c) Three gear locking coefficients $L_{12}$ and $L_{13}$ as a function of external torque with center-of-mass distance $d_{CM}=1.67$ nm between neighboring gears. (d) Three gears locking coefficient from the analytic two-gear potential for different barrier heights $V_1=0$, $70$ meV and $140$ meV.  
}
  \label{Fig:Fig6}
\end{figure*}
In order to see if coupled gears can rotate together, we apply now a constant external torque to one gear (e.g.\ the left gear in Fig.~\ref{Fig:Fig4}\ (a)). To quantify the collective rotation, we define the locking coefficient:
\begin{equation}\label{eq:locking_coeff}
    L_{12}=\frac{1}{T_s}\int_0^{T_s} \frac{\theta_2(t)}{\theta_1(t)}\ dt
\end{equation}
with simulation time $T_s=500$ ps. One can easily see, that $L_{12}=-1$ corresponds to the situation when the gears are perfectly interlocked ($\theta_2\approx -\theta_1$ for all times). On the contrary, if two gears are not interlocked then $L_{12}$ approaches zero (only $\theta_1$ is increasing and $\theta_2\approx 0$). In Fig.~\ref{Fig:Fig6}\ (a), we plot the locking coefficient as a function of external torque $\tau_{ext}$ for several center-of-mass distances $d_{CM}$, and we find that there are three possible qualitatively different scenarios.

First, for $d_{CM}=1.67$ nm, if we gradually ramp up the external torque from 0 to 1.6 nN$\cdot$\AA, $L_{12}$ is non-vanishing but smaller than -1. In fact, the molecules under small external torque are barely rotated. We denote this regime  the \textit{underdriving phase}. Secondly, when $1.6$ nN$\cdot$\AA $<\tau_{ext}<2$ nN$\cdot$\AA, one can clearly see that there is an abrupt jump followed by a plateau with $L_{12}=-1$, which means the two gears suddenly show interlocked rotations. We thus call this region  the  \textit{driving phase}. Finally, if we increase the torque even further ($\tau_{ext}>2$ nN$\cdot$\AA), then $L_{12}$ decays to zero, since the first gear has been driven too strongly and  the second gear cannot follow the first one -- these regime is denoted as \textit{overdriving phase}. 

We have also tuned the center-of-mass distance $d_{CM}$ to $1.57$ and $1.74$ nm. As one can see, reducing $d_{CM}$ to $1.57$ nm shows that the locking coefficient is still very similar to $d_{CM}=1.67$ nm, since two gears are still coupled well. However, if we raise $d_{CM}$ to $1.74$ nm, then the interlocking plateau vanishes, meaning that at this distance it is not possible for the gears to have collective rotation under external driving, since two gears are already too far away to interlock.

\begin{figure}[t]\centering
 \includegraphics[width=0.5\textwidth]{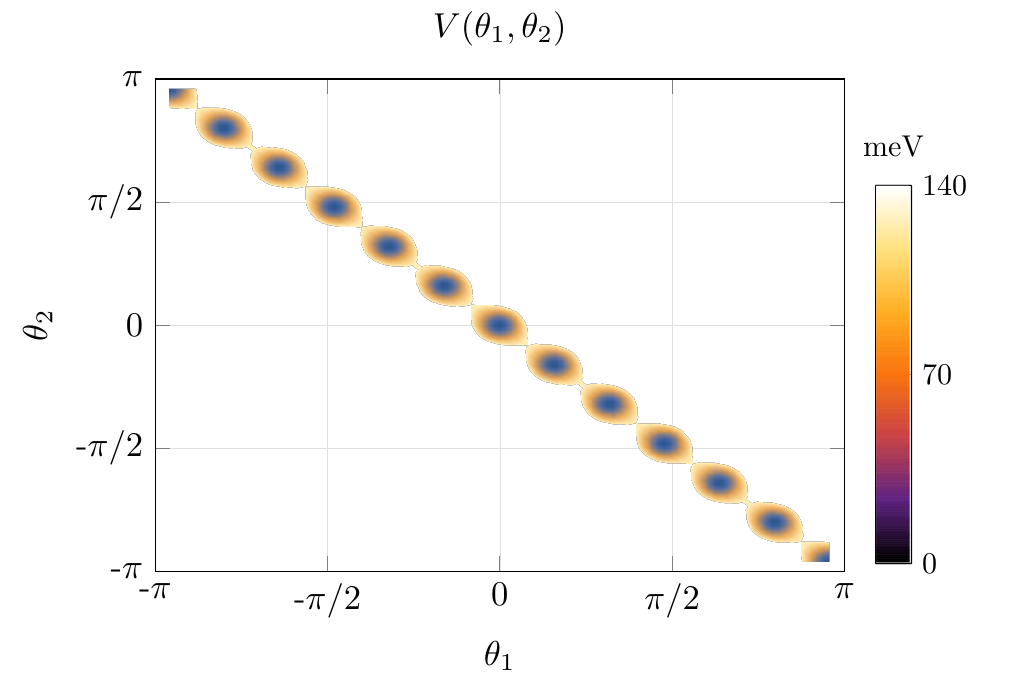}
 \caption{Analytic modelling of approximate two-gear interaction potential $V_{12}(\theta_1,\theta_2)$ with $V_0=450$ meV, $V_1=140$ meV and $k=50$.
}
\label{Fig:Fig7}
\end{figure}

{To understand how the different driving phases emerge, we propose an simplified two-gears interaction potential} given by the following expression:
\begin{eqnarray}
    V_{12}(\theta_1,\theta_2)&=&V_0+\left(1-\tanh{\left[k(1-\cos{(\theta_1+\theta_2)/3})\right]}\right)\times\nonumber \\ &&\left(-V_0+V_1\sin^2{\left[3(\theta_1-\theta_2)\right]}\right)
\end{eqnarray}
with free parameters $V_0$, $V_1$ and $k$. From this potential, one can plot the profile as shown in Fig.~\ref{Fig:Fig7} and see that there are 12 minima along the diagonal as motivated by the previous analysis of the MD simulation results. The inner barrier height between neighboring minima is defined by $V_1$, which is related to the threshold for the onset of the driving phase. On the other hand, the overall barrier perpendicular to the diagonal is controlled by $V_0$, which corresponds to the threshold for the onset of the overdriving phase. Additionally, we need a parameter $k$ to describe the well-width perpendicular to the diagonal. 
If $V_1$ is vanishing, we recover Eq.\ (\ref{eq:V_ideal}) which entails the total angular momentum conservation.

Using this potential energy ansatz, We can compute the locking coefficient  $L_{12}$ by solving the following coupled classical equations of motion:
\begin{subequations}
\begin{eqnarray}
    I_1\ddot{\theta_1}&=&-\gamma \dot{\theta_1}-\frac{\partial V_{12}}{\partial \theta_1}+\tau_{ext}\;,\\
    I_2\ddot{\theta_2}&=&-\gamma \dot{\theta_2}-\frac{\partial V_{12}}{\partial \theta_2}
\end{eqnarray}
\end{subequations}
with all the other parameters being the same as in the MD simulation. In Fig.~\ref{Fig:Fig6}\ (b), we use the parameters $V_0=450$ meV and $k=50$ to calculate $L_{12}$. For $V_1=0$, there is no underdriving phase and the two gears rotate even for very small external torque. However, one still observes the overdriving phase, since we have finite $V_0$, as long as $\tau_{ext}$ is large enough to break the interlocked gears. For $V_1=140$ meV one qualitatively reproduces the locking coefficient as observed in the MD simulation. During the rotation, two gears will collectively see a periodic potential as shown on the diagonal of Fig.~\ref{Fig:Fig7}, which means the center-of-mass kinetic energy will be reduced when crossing barriers. Therefore, the energy will be transferred to other degrees of freedom, like internal motion or gear deformation. This shows that when two coupled gears are trying to rotate, it is necessary for them to deform (depending on the stiffness of the molecule) in order to adopt a new conformation. Therefore, one can imagine that, when the barrier height $V_1$ reduces to $70$ meV, the molecule becomes stiffer and in turn it is easier to have collective rotations and display a wider interlocking plateau.

\subsubsection{Three gears system}
So far, we have analysed the conditions under which two gears can undergo collective rotation under the action of an external torque. One might ask how the situation changes for longer gear trains. We therefore consider at this point the case of three interacting gears, whose rotational dynamics satisfies the following set of equations: 
\begin{subequations}
\begin{eqnarray}
    I_1\ddot{\theta_1}&=&-\gamma \dot{\theta_1}-\frac{\partial V_{12}}{\partial \theta_1}+\tau_{ext}\;,\\
    I_2\ddot{\theta_2}&=&-\gamma \dot{\theta_2}-\frac{\partial V_{12}}{\partial \theta_2}-\frac{\partial V_{23}}{\partial \theta_2}\;,\\
    I_3\ddot{\theta_3}&=&-\gamma \dot{\theta_3}-\frac{\partial V_{23}}{\partial \theta_3}\;.
\end{eqnarray}
\end{subequations}
The calculated locking coefficients -- since there are three gears, we have two locking coefficients $L_{12}$ and $L_{13}$ -- are shown in  Fig.~\ref{Fig:Fig6}\ (c) for $1.67$ nm separation. Note that $L_{13}$ has to be positive (first and third gear should rotate in the same direction) and $L_{12}$ is negative as before. For $L_{13}$ and $L_{12}$ we do not see any interlocking plateaus. This implies that when applying a torque, we have either no rotation or overdriving, since the gears are too soft to be interlocked. Similarly, in the analytic calculation (see Fig.~\ref{Fig:Fig6}\ (d)), the width of the plateau has narrowed compared to the two gears case. For $V_1=140$meV, the plateau is also vanishing, which is consistent with the MD simulation results.

\section{\label{sec:Conclusion}Summary and Discussion}
By relating atomistic MD simulations to semi-analytical models, we have studied the rotational behavior of up to three gears made of hexa (4-tert-butylphenyl) benzene in thermal equilibrium as well as under an externally applied torque. A single gear behaves like a Brownian rotor in thermal equilibrium, displaying a linear increase of the ensemble average of its angle-variance with time. Under external torque, results from model calculations and MD simulations are also in good agreement. 
For two gears,  the interaction potential, displaying a set of $12$ metastable states  corresponding to either parallel or anti-parallel conformations of the tert-butyl groups, was extracted from the MD simulation data. In the presence of an external torque,  two gears can have collective rotations, but three gears did not display a concerted motion due to the stiffness of the molecules. Moreover, we have shown that the collective rotation is sensitive to the center-of-mass distance of the gears. Depending on the magnitude of the applied torque, we introduced a classification of the motion into underdriving, driving and overdriving phases, which were associated  to no collective rotations, collective rotations, and only single gear rotations, respectively.

Experimentally, driving the train of gears can be performed either mechanically or by inelastic excitation.\cite{Moresco2015} In the former case, rotations are considered to be induced by pushing one leg of the molecule gear using an STM tip. Clearly, in this situation the induced effective torque depends crucially on the flexibility of the molecule. If the latter is too soft, the torque due to the tip cannot be efficiently transferred to the ``driving'' gear and it might be difficult to induce collective motion of the whole train $-$ as we have shown in Sec.\ \ref{sec:drv_gears}. In computational approaches, the mechanical driving described above is often mimicked by predefining the rotation angle of the first gear and subsequently minimizing the energy with respect to the other gears\cite{Hove2018,Hove2018a}. In this minimization approach, however, the initial torque-transfer is not considered. Instead, an optimal path in the free-energy landscape is found, which might not be accessible for the gears. In this respect, the driving approach used in this work can be considered as complementary. In order to conclusively predict if collective motion can be observed in a train of gears, the driving mechanism (e.g.\ the STM tip) has to be included in the modelling.

\begin{acknowledgments}
We would like to thank A.~Kutscher, A.~Mendez, A.~Raptakis, T.~K{\"u}hne, R.~Biele, A.~Dianat and F.~Moresco for very useful discussions and suggestions. This work has been supported by the International Max Planck Research School (IMPRS) for ``Many-Particle Systems in Structured Environments'' and also by the European Union Horizon 2020 FET Open project "Mechanics with Molecules" (MEMO, grant nr.\ 766864).
\end{acknowledgments}

\bibliography{paper}
\end{document}